\documentclass[12pt]{article}
\pdfoutput=1

\usepackage{amsmath}
\usepackage{amsfonts}
\usepackage{amscd}
\usepackage{amsthm}
\usepackage{setspace}

\usepackage{graphicx}
\usepackage{authblk}
\usepackage{caption}
\usepackage{ytableau}
\usepackage{mathtools}

\usepackage[OT2,T1]{fontenc}
\DeclareSymbolFont{cyrletters}{OT2}{wncyr}{m}{n}
\DeclareMathSymbol{\Sha}{\mathalpha}{cyrletters}{"58}

\def\Z{\mathbb{Z}}

\def\P{\mathbb{P}}

\def\til{\tilde}

\begin{document}

\begin{titlepage}

\begin{flushright}
KEK-TH 2057
\end{flushright}

\vskip 1cm

\begin{center}

{\large $SU(n) \times \Z_2$ in F-theory on K3 surfaces without section as double covers of Halphen surfaces}

\vskip 1.2cm

Yusuke Kimura$^1$ 
\vskip 0.4cm
{\it $^1$KEK Theory Center, Institute of Particle and Nuclear Studies, KEK, \\ 1-1 Oho, Tsukuba, Ibaraki 305-0801, Japan}
\vskip 0.4cm
E-mail: kimurayu@post.kek.jp

\vskip 1.5cm
\abstract{We investigate F-theory models with a discrete $\Z_2$ gauge symmetry and $SU(n)$ gauge symmetries. We utilize a class of rational elliptic surfaces lacking a global section, known as Halphen surfaces of index 2, to yield genus-one fibered K3 surfaces with a bisection, but lacking a global section. We consider F-theory compactifications on these K3 surfaces times a K3 surface to build such models. We construct Halphen surfaces of index 2 with type $I_n$ fibers, and we take double covers of these surfaces to obtain K3 surfaces without a section with two type $I_n$ fibers, and K3 surfaces without a section with a type $I_{2n}$ fiber. We study these models to advance the understanding of gauge groups that form in F-theory compactifications on the moduli of bisection geometries. 
\par Our results also show that the Halphen surfaces of index 2 can have type $I_n$ fibers up to $I_9$. We construct an example of such a surface and determine the complex structure of the Jacobian of this surface. This allows us to precisely determine the non-Abelian gauge groups that arise in F-theory compactifications on genus-one fibered K3 surfaces obtained as double covers of this Halphen surface of index 2, with a type $I_9$ fiber times a K3 surface. We also determine the $U(1)$ gauge symmetries for compactifications when K3 surfaces as double covers of Halphen surfaces with type $I_9$ fiber are ramified over a smooth fiber.}  

\end{center}
\end{titlepage}

\tableofcontents
\section{Introduction}
\par Building models in particle physics using the F-theory \cite{Vaf, MV1, MV2} approach has several advantages; in this approach, the $SU(5)$ GUT model is naturally realized with matter fields in $SO(10)$ spinor representation. Additionally, the F-theory approach can avoid the problem of weakly coupled heterotic string theory, addressed in \cite{Witten}. Furthermore, up-type Yukawa couplings can be generated in this approach without difficulty. Local models \cite{DWmodel, BHV1, BHV2, DWGUT} have been mainly considered in recent studies of F-theory. Gravity, however, decouples when the local models are considered. Therefore, global models need to be investigated to discuss the problems related to gravity, such as the inflation. In this note, we analyze the global geometric structures of the compactification spaces in F-theory. 
\par Compactification spaces in F-theory have the structure of a genus-one fibration. The complex structure of a torus as a fiber of a genus-one fibration of a compactification space is identified with the axio-dilaton, enabling the axio-dilaton to possess $SL(2,\Z)$ monodromy. 7-branes in F-theory are wrapped on the components of the locus in the base space of a genus-one fibration, over which the fiber degenerates and becomes singular, namely {\it the discriminant locus}. The gauge symmetries and matter that arise are determined from the structure of the genus-one fibration of the compactification space. \cite{Kod1, Kod2} classified the types of singular fibers of elliptic surfaces \footnote{\cite{Ner, Tate} discussed techniques to determine the type of singular fibers of elliptic surfaces.}. The types of the singular fibers of a genus-one fibration correspond to the non-Abelian gauge groups that arise on the 7-branes in F-theory compactification \cite{MV2, BIKMSV}. This relationship is summarized in Table \ref{table:singularityand fibertype} below as the correspondence of the singularity types of the compactification space and the types of the singular fibers of a genus-one fibration that the compactification space admits. 

\begingroup
\renewcommand{\arraystretch}{1.1}
\begin{table}[htb]
\begin{center}
  \begin{tabular}{|c|c|} \hline
$\begin{array}{c}
\mbox{Type of} \\
\mbox{singular fiber}
\end{array}$ & Singularity type \\ \hline
$I_n$ ($n\ge 2$) & $A_{n-1}$ \\
$I^*_m$ ($m\ge 0$) & $D_{4+m}$ \\
$III$ & $A_1$ \\
$IV$ & $A_2$ \\
$IV^*$ & $E_6$ \\ 
$III^*$ & $E_7$ \\
$II^*$ & $E_8$ \\
$I_1$ & none. \\
$II$ & none. \\ \hline
\end{tabular}
\caption{The types of the singular fibers and the corresponding singularity types of the compactification space.}
\label{table:singularityand fibertype}
\end{center}
\end{table}  
\endgroup 

\par Some genus-one fibrations admit a global section, while the others do not. Elliptic fibrations with a global section have been studied in F-theory compactifications \cite{MorrisonPark, MPW, BGK, BMPWsection, CKP, AL, LSW, KSW2015, MP2, BMW2017, KimuraMizoguchi, Kimura1802, LRW2018}. In recent studies, F-theory compactifications on genus-one fibrations without a global section were analyzed. See, for example, \cite{BM, MTsection, AGGK, KMOPR, GGK, MPTW, MPTW2, CDKPP, LMTW, K, K2, KCY4, Kdisc, Kimura1801} for studies on F-theory compactifications on genus-one fibrations without a section. One of the reasons that the models that lack a section are considered is that a discrete gauge symmetry \footnote{See, e.g., \cite{KNPRR, ACKO, BS, HSsums, CIM, BCMRU, BCMU, HS, NRRV2013, BRU, KKLM, BGKintfiber, HS2, GPR, CGP} for recent studies of discrete gauge symmetries.} arises in these models. The mechanism which accounts for the origin of discrete gauge symmetry in the moduli of F-theory on genus-one fibrations is discussed in \cite{MTsection}. The Tate--Shafarevich group of (the Jacobian fibration of) a genus-one fibration and the discrete gauge symmetry that arises in the model on the genus-one fibration are identified \cite{BDHKMMS}. When a genus-one fibered Calabi--Yau manifold $M$ is given, the Jacobian fibration $J(M)$ of it is considered. A genus-one fibered Calabi--Yau manifold $M$ and the Jacobian fibration $J(M)$ have the identical $\tau$ functions. The Calabi--Yau genus-one fibrations, the Jacobian fibrations of which are isomorphic to $J(M)$, form a group. This group is referred to as the Tate--Shafarevich group $\Sha(J(M))$. A discrete gauge group that arises in F-theory compactification on the Calabi--Yau genus-one fibration $M$ is given by the Tate--Shafarevich group $\Sha(J(M))$. 
\par In general, a discrete $\Z_n$ symmetry arises in F-theory compactification on a genus-one fibration which possesses an $n$-section. F-theory models with discrete gauge symmetries are discussed, for example, in \cite{MTsection, KMOPR, GGK, MPTW, MPTW2, BGKintfiber, CDKPP, LMTW, Kdisc, Kimura1801}. Discrete $\Z_2,\Z_3,\Z_4,\Z_5$ symmetries are mainly studied in these constructions. 
\par The structure of the genus-one fibration, including a multisection that it contains, needs to be analyzed to deduce gauge groups and matters that arise in F-theory compactifications. The demonstration of the existence of a model in F-theory with a discrete gauge symmetry with a specific gauge group is non-trivial. We show the existence of some models with a discrete $\Z_2$ symmetry with specific gauge groups in the moduli of F-theory.
\par In this note, we construct several models of F-theory compactifications with a discrete $\Z_2$ symmetry with type $I_n$ fibers. We advance the understanding of models with type $I_n$ fibers in the moduli of F-theory compactified on bisection geometries. Concretely, we first construct surfaces that belong to a class of rational elliptic surfaces without a global section, known as Halphen surfaces of index 2. We explicitly construct examples of these surfaces with type $I_4$, $I_7$, $I_8$, and $I_9$ fibers. By utilizing these surfaces, we yield K3 surfaces that lack a global section but have a bisection. F-theory compactification on the resulting K3 surfaces times a K3 surface show the existence of models with $SU(n)^2 \times \Z_2$ and $SU(2n) \times \Z_2$ gauge groups, $n=4,7,8,9$. This result can advance the understanding of the non-Abelian gauge symmetries that arise in the moduli of F-theory with a discrete $\Z_2$ gauge symmetry on bisection geometries. 
\par Rational elliptic surfaces admit a genus-one fibration, but do not necessarily have a global section. Halphen surfaces \footnote{Discussions on the structure of Halphen surfaces can be found in \cite{DolgachevZhang, CantatDolgachev}. An application of Halphen surfaces of index 2 to string theory is discussed in \cite{Kimura1801}.} are examples of rational elliptic surfaces that lack a global section. Halphen surfaces of index 2 can be constructed by blowing up $\P^1\times\P^1$ at 8 points of the specific configuration: We consider a bi-degree (4,4) curve in $\P^1\times\P^1$, $k=0$, with eight simple singularities (the curve $k=0$ may have more singularities, other than the eight singularities). We choose a smooth bi-degree (2,2) curve, $l=0$, which passes through these eight simple singularities. The process of blowing up these eight singular points yields a Halphen surface. The projection onto $\P^1$ induced by taking the ratio $[k:l^2]$ yields a genus-one fibration. The exceptional divisors that arise when the eight singularities are blown up yield bisections to the fibration. The structure of the Halphen surfaces of index 2 is reviewed in section \ref{ssec 2.1}. It is explained there that Halphen surfaces do not have a global section.
\par In this study, we construct several examples of Halphen surfaces of index 2 with type $I_n$ fibers. We mainly consider the blow-ups of $\P^1\times\P^1$ at 8 points to construct Halphen surfaces of index 2. These constructions yield Halphen surfaces of index 2 with $I_4$, $I_7$, and $I_8$ fibers. We also consider a blow-up of $\P^2$ at 9 points to yield a Halphen surface of index 2 with an $I_9$ fiber. We consider the case in which the polynomial $k=0$ is reducible into lines or curves to construct these surfaces. Specific configurations of these lines and curves yield type $I_n$ fibers \footnote{Halphen surfaces of index 2 with type $I_m$ fibers, $m=2,3,5,6$, are constructed in \cite{Kimura1801}.} after blow-ups, $n=4,7,8,9$. 
\par Taking double covers of the examples of the Halphen surfaces of index 2 that we construct in this study yields genus-one fibered K3 surfaces \footnote{The K3 surfaces with involution are considered in \cite{Nikfactor}. These K3 surfaces belong to this class.} which lack a global section \footnote{Similar constructions obtained by considering double covers of Halphen surfaces of index 2, realized as blow-ups of $\P^2$ at nine points can be found in \cite{Kimura1801}.}. The resulting K3 surfaces are bisection geometries. F-theory compactifications on these K3 surfaces times a K3 surface reveal the existence of models in which a discrete $\Z_2$ symmetry and $SU(N)$ gauge group arise.
\par We will show in section \ref{ssec 2.4} that when a Halphen surface of index 2 possesses a type $I_n$ fiber, the upper bound on the degree $n$ is 9. The Halphen surface of index 2 with a type $I_9$ fiber that we construct in this study provides such an example. This fact imposes some constraints on the non-Abelian gauge symmetries that arise in F-theory compactification on genus-one fibered K3 surfaces constructed as double covers of Halphen surfaces with type $I_n$ fibers times a K3 surface. 
\par This study is structured as follows: In section \ref{sec 2}, after we briefly review the structure of Halphen surfaces of index 2, we construct Halphen surfaces of index 2 with type $I_n$ fibers, $n=4,7,8,9$. The constructions of these surfaces are discussed in section \ref{ssec 2.2} and section \ref{ssec 2.3}. We also show in section \ref{ssec 2.4} that when a Halphen surface of index 2 has an $A_n$ singularity, the highest is $A_8$. Thus, a Halphen surface of index 2 can have a type $I_m$ fiber up to an $I_9$ fiber. The Tate--Shafarevich groups of the Jacobian fibrations of the Halphen surfaces of index 2 are trivial. For the Halphen surfaces, the information of the multisections that they possess is contained in the Weil--Ch\^{a}telet groups of their Jacobians. We discuss this in section \ref{ssec 2.5}.
\par In section \ref{sec 3}, we construct genus-one fibered K3 surfaces lacking a section using examples of Halphen surfaces of index 2 as described in section \ref{sec 2}. There are two types of constructions of K3 surfaces. These two types of constructions consider taking double covers of Halphen surfaces that are obtained in section \ref{ssec 2.2} and section \ref{ssec 2.3}, ramified over either a smooth fiber, or a singular fiber. Consequently, we obtain two types of K3 surfaces without a section. The two constructions yield K3 surfaces with type $I_n$ fibers, and K3 surfaces with type $I_{2n}$ fibers. These are bisection geometries. The two constructions are presented in section \ref{ssec 3.1}. In section \ref{ssec 3.2} and section \ref{ssec 3.3}, we discuss F-theory compactifications on the genus-one fibered K3 surfaces lacking a section as described in section \ref{ssec 3.1} times a K3 surface. This yields a four-dimensional theory. We study the gauge symmetries that arise in these compactifications. Because the constructed K3 surfaces are bisection geometries, discrete $\Z_2$ gauge symmetries arise in these models. 
\par We discuss the Jacobian fibrations \footnote{Discussion of the Jacobian fibrations of elliptic curves can be found in \cite{Cas}.} of Halphen surfaces of index 2, and the constructed K3 surfaces as described in section \ref{sec 3} in section \ref{sec 4}. These surfaces have bisection geometries, therefore, as discussed in \cite{BM}, the Jacobian fibrations of these surfaces always exist. By taking the Jacobian fibration, the double fiber of a Halphen surface of index 2 becomes a smooth fiber. In section \ref{ssec4.2}, we determine the Weierstrass equation of the Jacobian fibration of a K3 surface obtained as a double cover of the Halphen surface with a type $I_9$ fiber ramified over a smooth fiber. We also determine the Mordell--Weil rank of this Jacobian fibration. Using this result, we deduce that there is no $U(1)$ gauge symmetry in F-theory compactification on the direct product of a K3 surface obtained as a double cover of the Halphen surface with a type $I_9$ fiber, ramified over a smooth fiber, times a K3 surface. We state our concluding remarks in section \ref{sec 5}. 

\section{Halphen surfaces of index 2 with type $I_4, I_7, I_8, I_9$ fibers}
\label{sec 2}
\subsection{Review of Halphen surfaces of index 2 constructed as blow-ups of $\P^1\times\P^1$}
\label{ssec 2.1}
We review the structure of Halphen surfaces of index 2 that are constructed as a blow-up of $\P^1\times\P^1$ at eight points. These surfaces are genus-one fibered and have a bisection, but they lack a global section. 
\par We take a curve, $k=0$, of bi-degree (4,4) in $\P^1\times\P^1$ that has 8 simple singularities. (The curve $k=0$ can have more singularities, other than the 8 singularities.) We choose a smooth bi-degree (2,2) curve, $l=0$, which passes through the 8 singularities of the curve $k=0$. The (2,2) curve in $\P^1\times\P^1$ has 9 monomials, therefore, a smooth (2,2) curve that passes through the fixed 8 points always exists. 
\par The blow-up of $\P^1\times\P^1$ at the 8 simple singularities of the (4,4) curve $k=0$ gives a rational surface. The ratio of the (4,4) curve $k=0$ to the square of the (2,2) curve $l=0$, $[k:l^2]$, gives a projection onto $\P^1$. This map endows the rational surface obtained as the blow-up of $\P^1\times\P^1$ at the 8 simple singularities of the curve $k=0$ with a fibration structure. The rational surface obtained as the blow-up of $\P^1\times\P^1$ at the 8 simple singularities together with the fibration structure induced from the map $[k:l^2]$, is called a Halphen surface of index 2 \footnote{Halphen surfaces of index 2, constructed as blow-ups of $\P^2$ at nine singularities are reviewed in \cite{Kimura1801}.}.
The curve $k=0$ yields a divisor belonging to the following complete linear system:
\begin{equation}
|4H_1+4H_2-2\Sigma^8_{i=1} P_i|,
\end{equation}
where we have used $H_1$ to denote the line class that the first $\P^1$ in the product $\P^1\times\P^1$ defines, and $H_2$ to denote the line class that the second $\P^1$ in $\P^1\times\P^1$ defines. We used $P_i$ to denote the exceptional divisors that arise when the 8 simple singularities are blown up. The bi-degree (2,2) curve $l=0$ defines a divisor which belongs to the following complete linear system:
\begin{equation}
|2H_1+2H_2-\Sigma_{i=1}^8 P_i|.
\end{equation}
\par A fiber $F$ of the projection $[k:l^2]$ onto $\P^1$ is given by
\begin{equation}
\label{genus-one fiber in 2.1}
k+\lambda\, l^2=0,
\end{equation}
where $\lambda$ denotes a constant. This is a (4,4) curve with eight simple singularities. Therefore, by the genus formula, the genus $g(F)$ of a fiber (\ref{genus-one fiber in 2.1}) is given by: 
\begin{equation}
g(F)=(4-1)(4-1)-\delta=9-8=1.
\end{equation}
$\delta=8$ is the number of singularities of a fiber (\ref{genus-one fiber in 2.1}). This shows that a generic fiber of the projection $[k:l^2]$ is a genus-one curve, namely, the projection $[k:l^2]$ is a genus-one fibration. 
\par The following equation describes the fiber over the point $[a:b]$ in the base $\P^1$:
\begin{equation}
a\cdot l^2-b\cdot k=0.
\end{equation}
The class $[k=0]$ represents the fiber at the origin $[0:1]$ of the base $\P^1$, and the class $[l^2=0] \sim 2\cdot [l=0]$ represents the fiber at the infinity $[1:0]$. Fibers of a rational surface are linearly equivalent. Thus we obtain the following linear equivalence relations: 
\begin{equation}
\label{equivalence relations in 2.1}
[k=0] \sim 2\cdot [l=0] \sim F.
\end{equation}
$F$ in (\ref{equivalence relations in 2.1}) denotes the fiber class. The linear equivalence relations (\ref{equivalence relations in 2.1}) means that the intersection number of any divisor with the fiber $F$ is a multiple of 2. A global section and the fiber $F$ have the intersection number 1. Thus, it follows that a Halphen surface of index 2 does not have a global section. 
\par Blowing up the 8 singularities of the curve $k=0$ yields bisections to the genus-one fibration of a Halphen surface of index 2. The fiber at the infinity $[1:0]$ of the base $\P^1$ is a unique double fiber. 
\par In section \ref{ssec 2.2} and section \ref{ssec 2.3}, we particularly consider the cases in which the (4,4) curve $k=0$ is reducible into curves of bi-degree (1,0), (0,1), or (1,1) to construct Halphen surfaces of index 2 with type $I_n$ fibers, $n=4,7,8,9$. 

\subsection{Construction of Halphen surfaces with type $I_4, I_7, I_8$ fibers}
\label{ssec 2.2}
We construct Halphen surfaces of index 2 with type $I_4$, $I_7$, and $I_8$ fibers. We consider a blow-up of $\P^1\times\P^1$ at 8 points, and we choose specific curves $k=0$ to realize these constructions.
\subsubsection{Halphen surface with $I_4$ fiber}
\begin{figure}
\begin{center}
\includegraphics[width=\linewidth,height=\textheight,keepaspectratio]{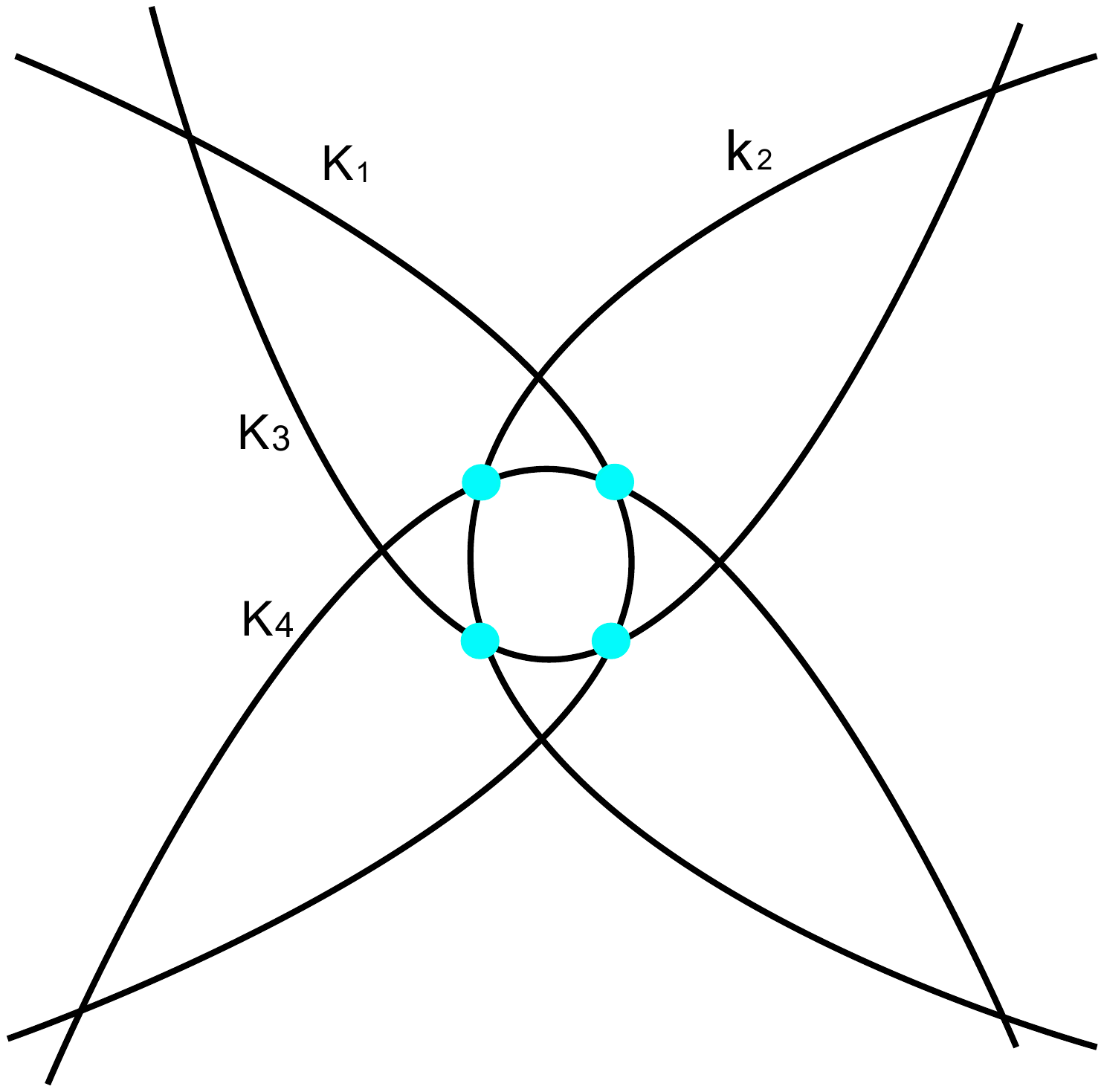}
\caption{\label{figI4}Configuration of the four (1,1) curves, $k_1, k_2, k_3, k_4$, and twelve intersection points. The blue dots represent the four points which are left un-blown up. The four (1,1) curves passing through the four points form a quadrangle. The remaining eight intersections are blown up.}
\end{center}
\end{figure}

We choose the bi-degree (4,4) polynomial $k$ as the product of four irreducible (1,1) curves $k_1, k_2, k_3$, and $k_4$:
\begin{equation}
k=k_1\, k_2\, k_3 \, k_4.
\end{equation}
Each pair of distinct (1,1) curves, $k_i$ and $k_j$, $i\ne j$, intersect at 2 points in $\P^1\times \P^1$ \footnote{Curves of bi-degrees $(a,b)$ and $(c,d)$ in $\P^1\times\P^1$ have the number of intersections as $ad+bc$.}. We have $\binom{4}{2}=6$ pairs of bi-degree (1,1) curves. Therefore, we have 12 intersection points in total. These are the simple singularities of the (4,4) curve $k=0$. We choose four points among these 12 intersection points so that the bi-degree (1,1) curves passing through these four points form a quadrangle. We show the image of four chosen points and the bi-degree (1,1) curves $k_i=0$, $i=1,2,3,4$ in Figure \ref{figI4}. The four chosen points are not blown up, and we consider the blow-up of the remaining eight points. Because each bi-degree (1,1) curve $k_i=0$, $i=1,2,3,4$, has the genus \footnote{A bi-degree ($a,b$) smooth curve in $\P^1\times\P^1$ has the genus $(a-1)(b-1)$.} $(1-1)(1-1)=0$, bi-degree (1,1) curve $k_i=0$ is rational, i.e, it is isomorphic to $\P^1$. Therefore, this construction yields a Halphen surface of index 2 with type $I_4$ fiber at the origin of the base $\P^1$ under the projection $[\Pi^4_{i=1}k_i : l^2]$. 

\subsubsection{Halphen surface with $I_7$ fiber}
\label{sssec 2.2.2}
\begin{figure}
\begin{center}
\includegraphics[width=\linewidth,height=\textheight,keepaspectratio]{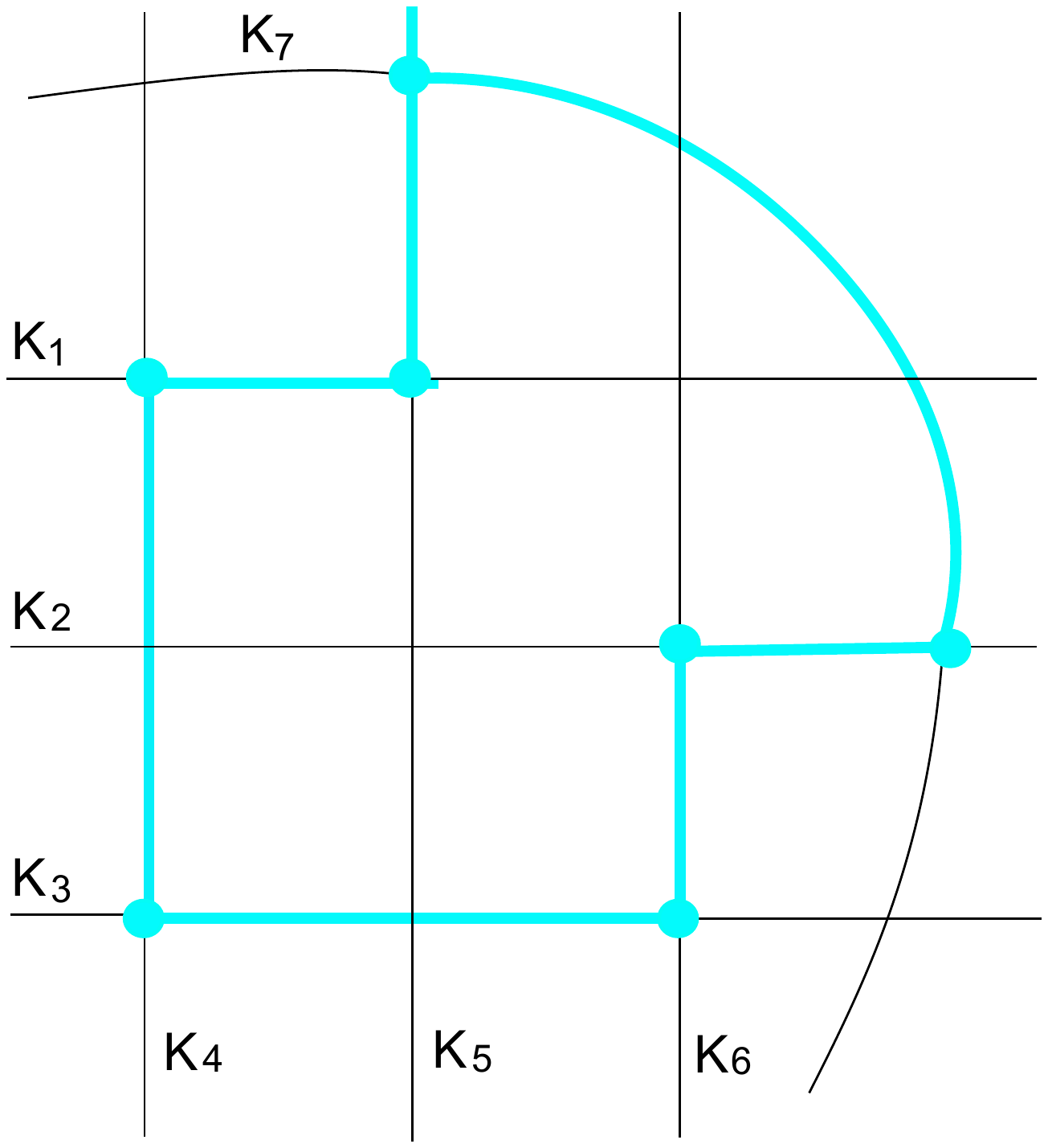}
\caption{\label{figI7}Configuration of the lines and the (1,1) curve, $k_1, \cdots, k_7$, and their fifteen intersections. The seven blue dots represent the seven points which are chosen to not undergo blowning up. The lines and the (1,1) curve passing through these chosen points form a 7-gon as indicated by the blue lines and the blue curve in the image.}
\end{center}
\end{figure}

We choose the bi-degree (4,4) curve $k$ as the product of three bi-degree (1,0) curves, $k_1, k_2, k_3$, three bi-degree (0,1) curves, $k_4, k_5, k_6$, and a bi-degree (1,1) curve $k_7$:
\begin{equation}
k = \Pi^7_{i=1} k_i.
\end{equation}
We assume that these seven curves, $k_1, \cdots, k_7$, are in a general position. A distinct pair of two bi-degree (1,0) curves do not intersect. Similarly, a distinct pair of two bi-degree (0,1) curves do not meet. A pair consisting of a (1,0) curve and a (0,1) curve meet at 1 point. There are 9 pairs of a (1,0) curve, $k_i$, $i=1,2,3$, and a (0,1) curve, $k_j$, $j=4,5,6$. The bi-degree (1,1) curve $k_7$ and a bi-degree (1,0) curve or (0,1) curve, $k_i$, $i=1, \cdots, 6$, intersect at 1 point. Therefore, we have $9+6=15$ intersection points in total. We choose seven points among these 15 intersection points, so that the curves $k_i$ passing through the seven chosen points form a 7-gon. The chosen seven points are not blown up, and we blow up the remaining 8 points. The image of the seven points and the seven irreducible curves, $k_1, \cdots, k_7$, are shown in Figure \ref{figI7}. Each of the seven irreducible curves, $k_1, \cdots, k_7$, has the genus 0. Therefore, they are each isomorphic to $\P^1$. This construction yields a Halphen surface of index 2, with a type $I_7$ fiber at the origin under the projection $[\Pi^7_{i=1}k_i : l^2]$. 

\subsubsection{Halphen surface with $I_8$ fiber}
\begin{figure}
\begin{center}
\includegraphics[width=\linewidth,height=\textheight,keepaspectratio]{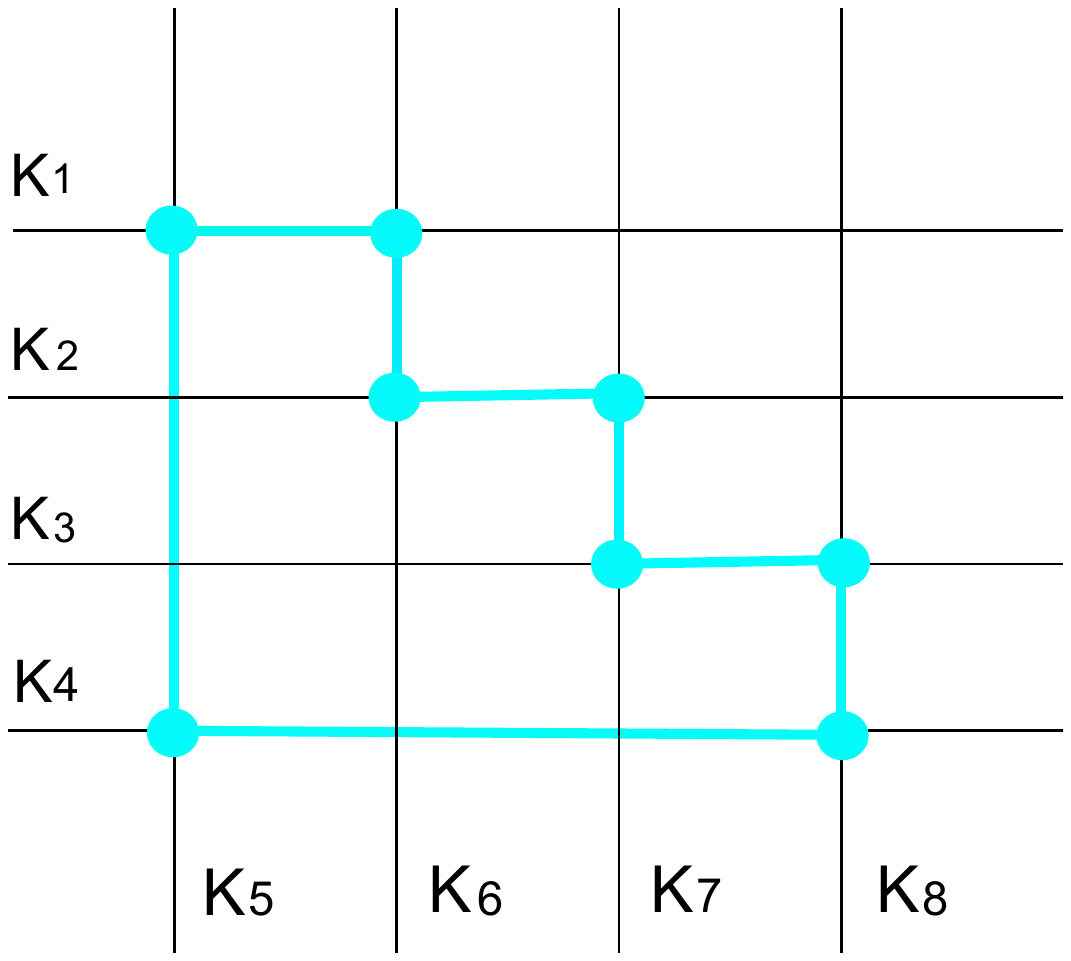}
\caption{\label{figI8}Configuration of the eight lines, $k_1, \cdots, k_8$. These lines have sixteen points of intersection in total. The eight blue dots represent the eight points which are chosen to remain un-blown up. As the blue lines in the image show, the lines passing through the chosen eight points form an octagon.}
\end{center}
\end{figure}

There are at least two ways to construct a Halphen surface of index 2 with a type $I_8$ fiber. One construction is given as follows: we choose the bi-degree (4,4) curve $k$ as the product of four bi-degree (1,0) curves, $k_1, \cdots, k_4$, and four bi-degree (0,1) curves, $k_5, \cdots, k_8$. These curves have $4 \times 4=16$ intersection points. We choose eight points among these 16 intersection points, so that the irreducible curves $k_i$, $i=1, \cdots, 8$, passing through the chosen eight points form an octagon. The chosen eight points are not blown up, and we blow up the remaining eight points. The image of the chosen eight points that were not blown up, and the eight irreducible curves $k_1, \cdots, k_8$, are shown in Figure \ref{figI8}. This construction yields a Halphen surface of index 2, with an $I_8$ fiber at the origin of the base $\P^1$ under the projection $[\Pi^8_{i=1}k_i : l^2]$.

\vspace{5mm}

\par
\begin{figure}
\begin{center}
\includegraphics[width=\linewidth,height=\textheight,keepaspectratio]{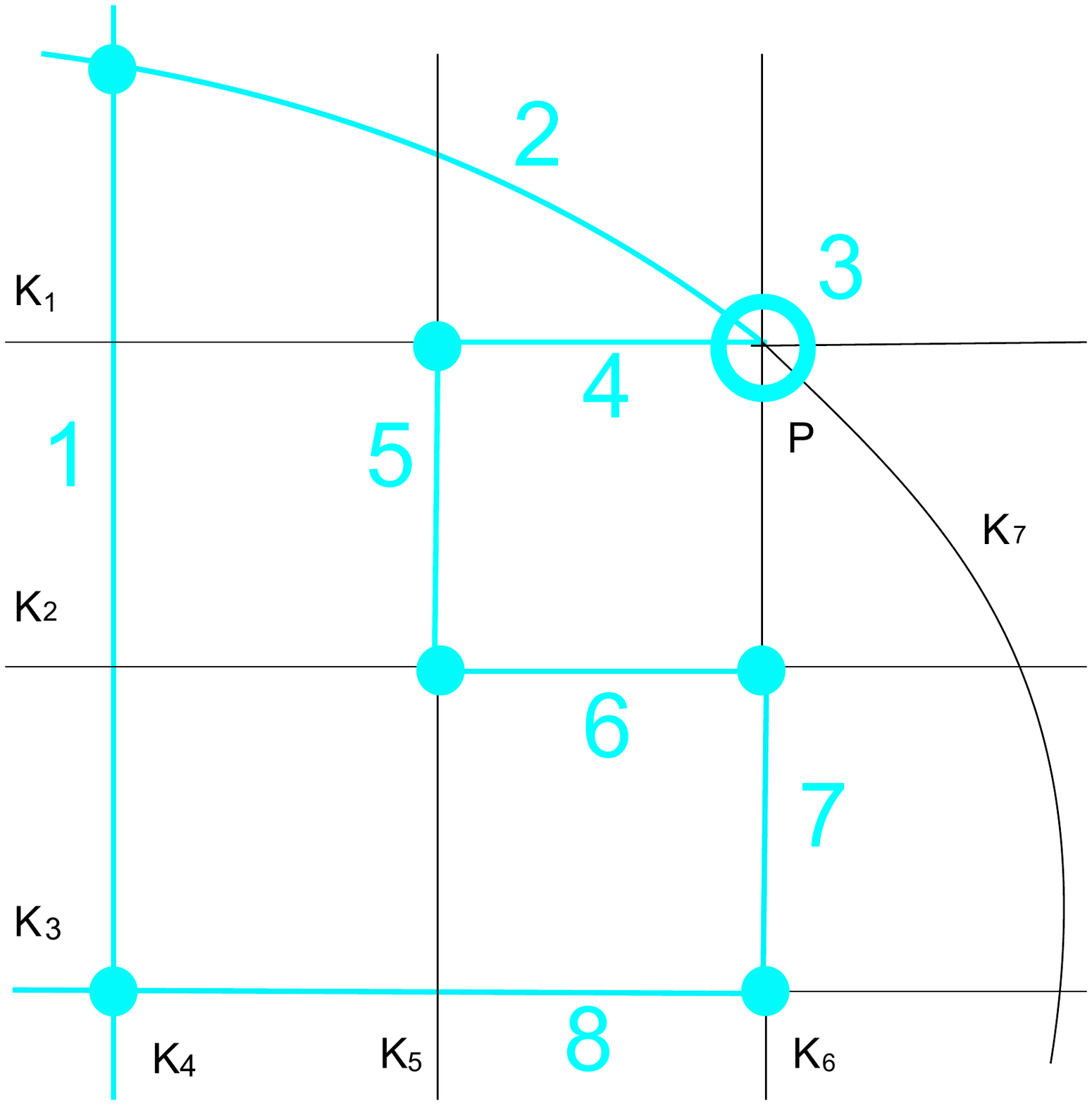}
\caption{\label{figI8another}The blue circle in the image numbered 3 represents the exceptional divisors $E_p$. We use the exceptional divisors $E_p$ as an edge to form an octagon. The curves used as the edges of the octagon are numbered with the blue numbers. We choose the two points, the intersections of $E_p$ with the curves $k_1$ and $k_7$, to not undergo blowning up. The six blue dots in the image represent the remaining six points that are chosen to remain un-blown up.}
\end{center}
\end{figure}

 Another construction of a Halphen surface of index 2 with a type $I_8$ fiber arises from the consideration of a special configuration of the seven irreducible curves $k_1, \cdots, k_7$, as described previously in section \ref{sssec 2.2.2}, to construct a Halphen surface of index 2 with a type $I_7$ fiber. We considered a general configuration of the seven curves $k_1, \cdots, k_7$, so that the bi-degree (1,1) curve meet with the six curves $k_1, \cdots, k_6$ at six points in the previous section \ref{sssec 2.2.2}. Here, we consider a special configuration in which the (1,1) curve $k_7$ passes through the intersection point of the line $k_1$ and the line $k_6$. We refer to this intersection point of the curves $k_1$, $k_6$, and $k_7$ as $p$. We then consider the blow-up of the intersection point $p$. This operation yields an exceptional divisor $E_p\cong \P^1$ at the point $p$, and this separates the three curves $k_1$, $k_6$, and $k_7$. Each of these curves intersects the exceptional divisor $E_p$ at 1 point. These three points together with the other 12 intersections of the curves give 15 intersection points in total. Eight specific points among these are chosen so that the curves passing through the chosen eight points form an octagon. The exceptional divisor $E_p$ was utilized as an edge of this octagon. The chosen eight points were not blown up, and we blew up the remaining seven points \footnote{We once blew up the point $p$. Therefore, the remaining number of blow-ups required to yield a Halphen surface is eight minus one, i.e., 7.}. The image of the configuration of the curves forming an octagon is shown in Figure \ref{figI8another}. This yields a Halphen surface of index 2 with a type $I_8$ fiber at the origin of the base $\P^1$ under the projection $[\Pi^7_{i=1}k_i : l^2]$.

\subsection{Construction of Halphen surface with type $I_9$ fiber}
\label{ssec 2.3}
As described in \cite{Kimura1801}, a Halphen surface of index 2 can also be constructed by taking a sextic curve $k$ in $\P^2$ with 9 simple singularities \footnote{As explained in \cite{Kimura1801}, the sextic curve $k$ can have more singularities, other than the 9 singularities.}, and by blowing up $\P^2$ at these 9 simple singularities. We choose a smooth cubic curve $l$ that passes through the 9 simple singularities of the curve $k$, and a genus-one fibration of the constructed Halphen surface is given by the projection $[k: l^2]$ onto $\P^1$. Using this type of construction of a Halphen surface of index 2, we yield a Halphen surface of index 2 with a type $I_9$ fiber \footnote{By considering the construction of a Halphen surface as the blow-up of $\P^2$ at nine simple singularities of a sextic curve, Halphen surfaces of index 2 with type $I_1, I_2, I_3, I_5, I_6$ fibers are constructed in \cite{Kimura1801}.}. 
\par
\begin{figure}
\begin{center}
\includegraphics[width=\linewidth,height=\textheight,keepaspectratio]{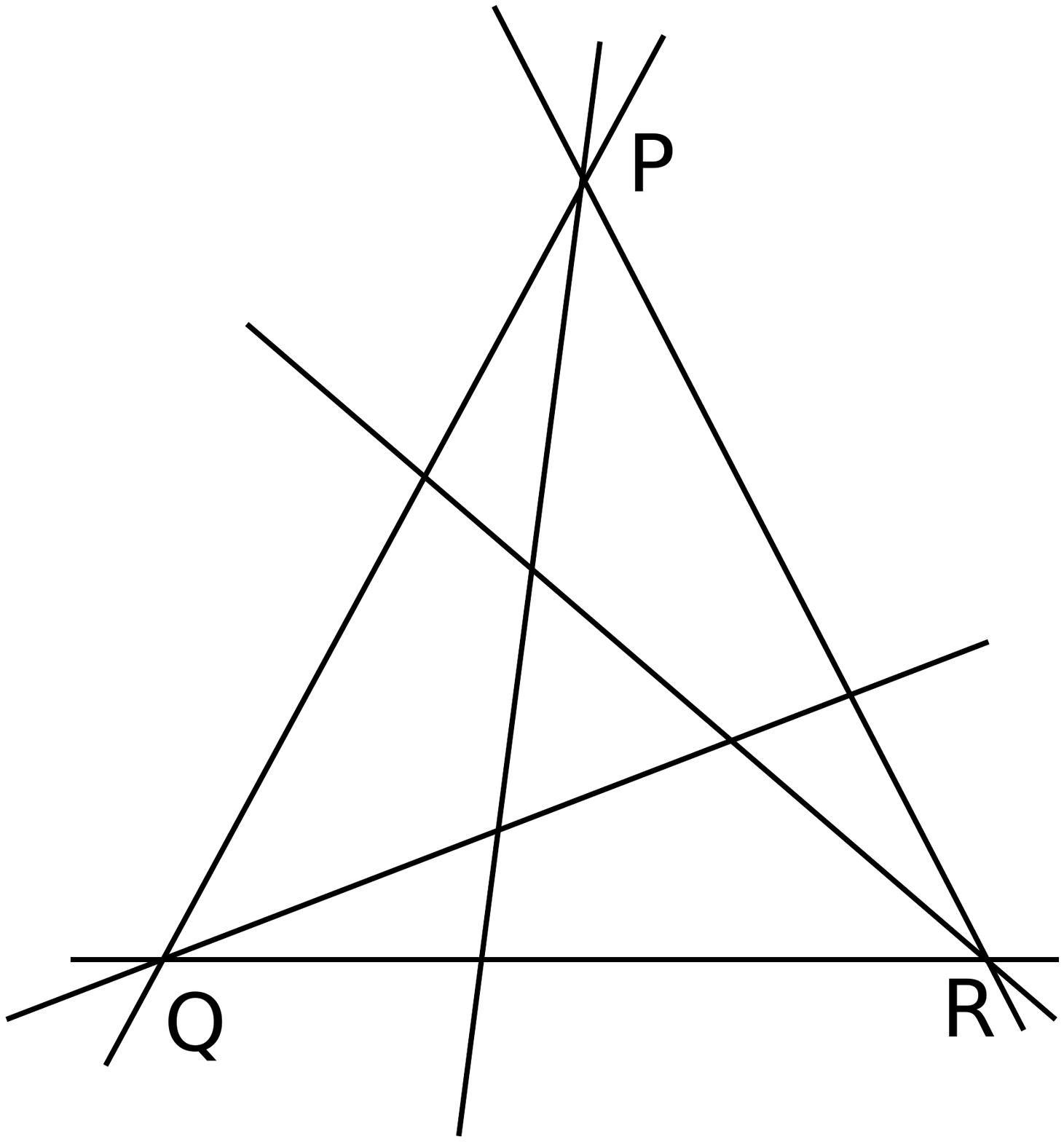}
\caption{\label{figI9}Configuration of the six lines, $k_1, \cdots, k_6$. Three lines meet at one point at each of the three points $P, Q,R$.}
\end{center}
\end{figure}

\begin{figure}
\begin{center}
\includegraphics[width=\linewidth,height=\textheight,keepaspectratio]{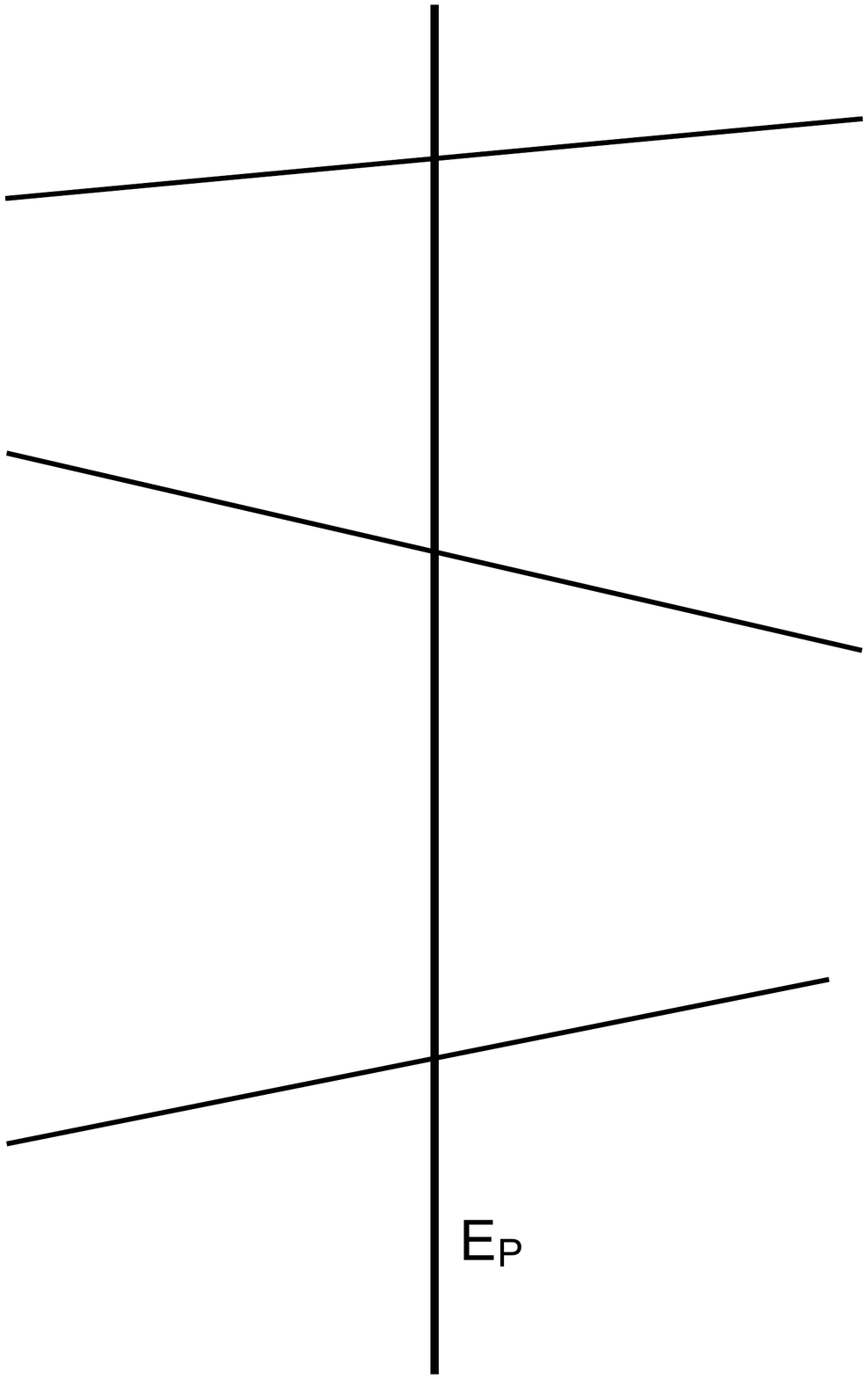}
\caption{\label{figblowup}Blow-up at $P$ separates the three lines that meet at one point at $P$. The vertical line represents the exceptional divisor $E_P$ that arises after blow-up. The horizontal three lines are the separated three lines. Each of these lines intersects the exceptional divisor at one point.}
\end{center}
\end{figure}

\begin{figure}
\begin{center}
\includegraphics[width=\linewidth,height=\textheight,keepaspectratio]{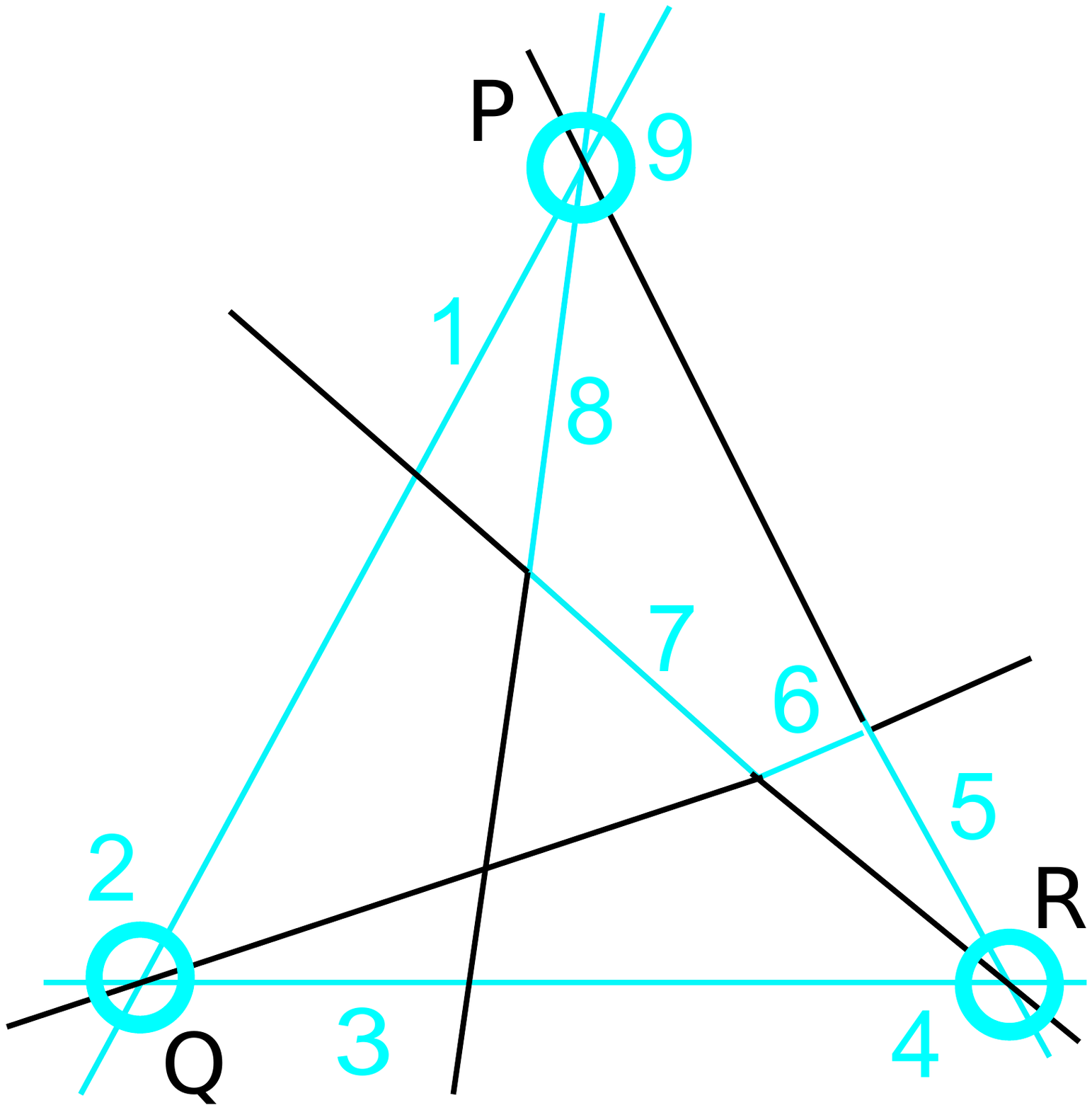}
\caption{\label{figI9blowup}The blue lines in the image are the edges of the 9-gon. Each of these edges are numbered, using the blue numbers. Blue circles numbered 2, 4, and 9 in the image represent the exceptional divisors $E_P, E_Q$, and $E_R$, respectively. These exceptional divisors are used as three edges, a part of the nine edges, to form the 9-gon.}
\end{center}
\end{figure}

 We consider the situation in which the sextic curve $k$ is reducible into six lines $k_i$, $i=1, \cdots, 6$:
\begin{equation}
k= \Pi^6_{i=1} k_i.
\end{equation}
We consider a specific configuration of the six lines $k_i$, $i=1, \cdots, 6$, as shown in Figure \ref{figI9}. At each of the three points $P, Q, R$ in Figure \ref{figI9}, three lines intersect at 1 point. Blowing up each of these points, $P, Q, R$, yields the exceptional divisors $E_P$, $E_Q$, and $E_R$, each of which is isomorphic to $\P^1$. This operation separates three lines that meet at 1 point. Each of these three lines meets an exceptional divisor at 1 point. We show the configuration of the three lines and the exceptional divisor $E_P$ when the blow-up is performed at the point $P$ in Figure \ref{figblowup}. The situations are identical for the points $Q, R$. After the three points $P, Q, R$ are blown up, there are 9 points for the intersections of the separated lines and the three exceptional divisors, $E_P$, $E_Q$, and $E_R$. Together with the other six intersections of the lines, there are 15 intersection points in total after the three points $P, Q, R$ are blown up. We chose nine specific points among the 15 intersection points so that the lines and the exceptional divisors passing through these chosen points form a 9-gon. The exceptional divisors $E_P$, $E_Q$, and $E_R$ were used as three edges among the nine edges forming this 9-gon. We show the configuration of the nine chosen points and lines forming the 9-gon in Figure \ref{figI9blowup}. The chosen nine points were not blown up, and we blew up the remaining six points. This yields a Halphen surface of index 2 \footnote{We blew up $\P^2$ at the three points, $P, Q, R$, therefore, the number of blow-ups remaining to yield a Halphen surface is nine minus three, i.e., six.} with a type $I_9$ fiber at the origin of the base $\P^1$ under the projection $[\Pi^6_{i=1}k_i : l^2]$.

\subsection{Upper bound on the degree of $A_n$ singularity of a Halphen surface of index 2, and the determination of the complex structure of the Jacobian fibration of a Halphen surface with $I_9$ fiber}
\label{ssec 2.4}
We constructed Halphen surfaces of index 2 with $A_n$ type singularities. By considering the Jacobian fibration, we show that a Halphen surface of index 2 with the singularity type $A_9$ or with $A_n$ type singularity of higher degree does not exist. Halphen surfaces of index 2 are bisection geometries. Therefore, as discussed in \cite{BM}, the Jacobian fibration of a Halphen surface of index 2 always exists. The Jacobian fibration of a Halphen surface is a rational elliptic surface with a global section. By the Shioda--Tate formula \cite{Shiodamodular, Tate1, Tate2}, the following equality \footnote{This equality is used to yield several families of rational elliptic surfaces with a section with various Mordell--Weil ranks in \cite{Kimura1802}. Identical pairs of these surfaces are glued to obtain elliptic K3 surfaces, on which $U(1)$ gauge symmetries of various ranks arise in F-theory compactifications, in \cite{Kimura1802}.} holds for the rank of the Mordell--Weil group \footnote{\cite{OS} classified the Mordell--Weil groups of rational elliptic surfaces that admit a global section.} and the rank of the singularity type of the Jacobian fibration of a Halphen surface:
\begin{equation}
rk\, {\rm MW}+ rk\, ADE =8,
\end{equation}
where $rk\, ADE$ denotes the rank of the singularity type of the Jacobian fibration of a Halphen surface. Particularly, the following inequality holds  
\begin{equation}
\label{inequality singularity in 2.4}
rk\, ADE \le 8.
\end{equation}
The singularity types of the original Halphen surface and the Jacobian fibration are identical. Therefore, the inequality (\ref{inequality singularity in 2.4}) shows that a Halphen surface of index 2 can have an $A_n$ type singularity up to $A_8$. This proves that a Halphen surface of index 2 can have a type $I_n$ fiber up to type $I_9$. 
\par A Halphen surface with a type $I_9$ fiber has the singularity type $A_8$, thus the Jacobian fibration of this Halphen surface is an extremal \footnote{Extremal rational elliptic surfaces are the rational elliptic surfaces with a section with the Mordell--Weil rank 0. They have the singularity types of rank 8.} rational elliptic surface \footnote{Applications of extremal rational elliptic surfaces that appear in the stable degeneration \cite{FMW, AM} of F-theory/heterotic duality \cite{Vaf, MV1, MV2, Sen, FMW} to string theory are discussed in \cite{Kimura1710}.}. The complex structures of the extremal rational elliptic surfaces were classified in \cite{MP}. The complex structure of an extremal rational elliptic surface with $A_8$ singularity is uniquely determined \cite{MP}. The extremal rational elliptic surface with $A_8$ singularity has 1 type $I_9$ fiber, and 3 type $I_1$ fibers \cite{MP}. This extremal rational elliptic surface is denoted as $X_{[9, 1, 1, 1]}$ in \cite{Kimura1710}. From the aforementioned argument, we deduce that the Jacobian fibration of the Halphen surface of index 2 with a type $I_9$ fiber constructed in section \ref{ssec 2.3} is isomorphic to the extremal rational elliptic surface $X_{[9, 1, 1, 1]}$. Utilizing this result, we precisely determine the non-Abelian gauge symmetry that arises in F-theory compactification on K3 surface obtained as a double cover of the Halphen surface of index 2 with a type $I_9$ fiber in section \ref{ssec 3.3}.

\subsection{The Tate-Shafarevich groups and the Weil-Ch\^{a}telet groups of the Jacobians of the Halphen surfaces}
\label{ssec 2.5}
\par The Jacobian fibrations of the Halphen surfaces generally have trivial Tate--Shafarevich groups. The Weil--Ch\^{a}telet group instead contains the information of the multisections for the Halphen surfaces. We discuss the structures of these groups of the Jacobian fibrations of the Halphen surfaces. 
\par The Tate--Shafarevich group is a subgroup of the Weil--Ch\^{a}telet group. Among genus-one fibrations as elements of the Weil--Ch\^{a}telet group, those which locally admit a section form a subgroup, and this subgroup is the Tate--Shafarevich group \cite{Cas}. 
\par A Halphen surface $X$ of index $n$ \footnote{Halphen surface of index $n$ is obtained by blowing up $\P^2$ at nine singularities of multiplicities $n$ of a degree $3n$ curve. The exceptional divisors yield $n$-sections.} does not have a global section, but it admits an $n$-section, therefore, the element $X$ generates a $\Z_n$ group in the Weil--Ch\^{a}telet group of the Jacobian fibration $J(X)$, $WC(J(X))$:
\begin{equation}
<X>\cong\Z_n\subset WC(J(X)).
\end{equation}
In this note, we constructed several examples of Halphen surfaces of index 2. These surfaces are bisection geometries, therefore, it follows from the aforementioned argument that the Weil--Ch\^{a}telet groups of the Jacobian fibrations of these surfaces contain $\Z_2$ groups.
\par We determined in section \ref{ssec 2.4} that the Jacobian fibration of the Halphen surface of index 2 with a type $I_9$ fiber as we constructed in section \ref{ssec 2.3} is isomorphic to the extremal rational elliptic surface $X_{[9, 1, 1, 1]}$. Thus, we find that the Weil--Ch\^{a}telet group $WC(X_{[9, 1, 1, 1]})$ of the extremal rational elliptic surface $X_{[9, 1, 1, 1]}$ contains a $\Z_2$ group, and the Tate--Shafarevich group $\Sha(X_{[9, 1, 1, 1]})$ is trivial:
\begin{equation}
WC(X_{[9, 1, 1, 1]})\supset\Z_2, \hspace{5mm} \Sha(X_{[9, 1, 1, 1]})\cong 0.
\end{equation}

\section{K3 surfaces without a section as double covers of Halphen surfaces, and gauge groups in F-theory compactifications}
\label{sec 3}
We utilize the Halphen surfaces of index 2 with type $I_n$ fibers that we constructed in section \ref{ssec 2.2} and section \ref{ssec 2.3} to yield a genus-one fibered K3 surface without a global section. As previously seen, the constructed Halphen surfaces have type $I_4, I_7, I_8, I_9$ fibers at the origin of the base $\P^1$. We consider two constructions of a genus-one fibered K3 surface without a section: double covers of Halphen surfaces ramified over a smooth fiber, and double covers of Halphen surfaces ramified over the singular fiber at the origin of the base $\P^1$. We consider F-theory compactifications on the resulting K3 surfaces without a global section times a K3 surface, and deduce gauge groups that arise. We also precisely determine the non-Abelian gauge symmetries for K3 surfaces obtained as double covers of Halphen surfaces of index 2 with type $I_9$ fibers.

\subsection{Constructions of K3 surfaces lacking a section as double covers of Halphen surfaces of index 2}
\label{ssec 3.1}
\subsubsection{Construction of K3 surfaces without a section as double covers of Halphen surfaces ramified along a smooth fiber}
\label{sssec 3.1.1}
A fiber of a Halphen surface of index 2 constructed as a blow-up of $\P^1\times\P^1$ at eight points is given by the following equation
\begin{equation}
\label{fiber equation in 3.1.1}
k +a\, l^2=0,
\end{equation}
where $k$ is a bi-degree (4,4) polynomial, so that the curve $k=0$ has eight simple singularities, and $l=0$ is a bi-degree (2,2) smooth curve that passes through these eight singularities, as described in section \ref{ssec 2.1}. $a$ is a constant.
\par We consider a double cover of a Halphen surface of index 2 ramified over a fiber 
\begin{equation}
\label{double cover K3 in 3.1.1}
\tau^2 = k +a\, l^2.
\end{equation}
The equation (\ref{double cover K3 in 3.1.1}) describes a double cover of a Halphen surface constructed as a blow-up of $\P^1\times\P^1$ at eight points, ramified over a bi-degree (4,4) curve. Thus, it gives a K3 surface.
\par Blowing up the simple singularities of the curve $k=0$ yields bisections to a Halphen surface of index 2. Using an argument similar to that presented in \cite{Kimura1801}, we deduce that the resulting K3 surface (\ref{double cover K3 in 3.1.1}) does not have a global section, and the pullback of a bisection to an original Halphen surface of index 2 yields a bisection to the resulting K3 surface (\ref{double cover K3 in 3.1.1}). 
\par We assume that 
\begin{equation}
a \ne 0
\end{equation}
in equation (\ref{double cover K3 in 3.1.1}). For generic values of $a\ne 0$, the equation (\ref{fiber equation in 3.1.1}) yields a smooth fiber. When $a$ takes the value 0, a fiber becomes singular, and the equation (\ref{fiber equation in 3.1.1}) describes a type $I_n$ fiber at the origin of the base $\P^1$. The case $a=0$ will be discussed in section \ref{sssec 3.1.2}. 
\par We find that the K3 surface (\ref{double cover K3 in 3.1.1}) that results as a double cover of a Halphen surface of index 2 is identical to the quadratic base change of the Halphen surface using an argument similar to that given in \cite{Kimura1801}. Therefore, the singular fibers that the K3 surface (\ref{double cover K3 in 3.1.1}) has are twice the number of the original Halphen surface, when the ramification locus of the double cover is a smooth fiber, namely $a \ne 0$. We discuss the case where the ramification locus of the double cover becomes a singular fiber, which occurs when $a=0$, in section \ref{sssec 3.1.2}. 
\par When $a=\infty$, the equation (\ref{double cover K3 in 3.1.1}) can be expressed as follows:
\begin{equation}
\tau^2 = l^2.
\end{equation}
This equation splits into the following two equations:
\begin{eqnarray}
\tau & = & l \\ \nonumber
\tau & = & -l.
\end{eqnarray}
This is the situation where a K3 surface degenerates into two rational elliptic surfaces as discussed in \cite{Kimura1710}. We assume that $a\ne\infty$ in this study.
\par A double cover of a Halphen surface of index 2 constructed as a blow-up of $\P^2$ at nine points, given by an equation of the same form as (\ref{double cover K3 in 3.1.1}), also gives a K3 surface without a section. This construction of a genus-one fibered K3 surface without a section is discussed in \cite{Kimura1801}.
\par The aforementioned argument shows that the K3 surface (\ref{double cover K3 in 3.1.1}) obtained as a double cover of a Halphen surface ramified along a smooth fiber has two type $I_n$ fibers when an original Halphen surface has a type $I_n$ fiber, $n=4,7,8,9$, as constructed in section \ref{ssec 2.2} and section \ref{ssec 2.3}. 

\subsubsection{Construction of K3 surfaces without a section as double covers of Halphen surfaces ramified along a singular fiber}
\label{sssec 3.1.2}
We discuss the case where $a$ in the equation (\ref{double cover K3 in 3.1.1}) takes the value 0, and a K3 surface as a double cover of a Halphen surface of index 2 is given by the following equation:
\begin{equation}
\label{double cover K3 in 3.1.2}
\tau^2 = k.
\end{equation}
For this case, the ramification locus of a K3 surface as a double cover (\ref{double cover K3 in 3.1.2}) occurs along the singular fiber
\begin{equation}
\label{singular fiber in 3.1.2}
k=0.
\end{equation}
The singular fiber (\ref{singular fiber in 3.1.2}) describes the type $I_n$ fiber of a Halphen surface at the origin of the base $\P^1$. 
\par An argument similar to that given in \cite{Kimura1801} proves that the K3 surface (\ref{double cover K3 in 3.1.2}) generically lacks a global section, but it has a bisection. 
\par The quadratic base change that corresponds to the double cover (\ref{double cover K3 in 3.1.2}) ramifies over type $I_n$ fiber at the origin of the base $\P^1$, and the resulting K3 surface as a double cover (\ref{double cover K3 in 3.1.2}) has a type $I_{2n}$ fiber \footnote{When the quadratic base change is ramified over a type $I_n$ fiber, two type $I_n$ fibers collide, instead of simply yielding two copies of type $I_n$ fiber, and they are enhanced to a singular fiber of type $I_{2n}$ \cite{SchShio}.}, instead of two $I_n$ fibers as described in section \ref{sssec 3.1.1}. 
\par When the simple singularities of the bi-degree (4,4) polynomial $k$ include a cusp, the pullback of the bisection which arises when the cusp is blown up splits into two sections \footnote{The curve $k+a\, l^2$ does not generally have a cusp for nonzero $a$.}. This is because a bisection that arises as an exceptional divisor when a cusp is blown up is tangent to the branching locus. The K3 surface (\ref{double cover K3 in 3.1.2}) admits a global section for this special situation. We do not consider this situation in this study, and we assume that the singularities of the polynomial $k$ do not have a cusp. 
\par The aforementioned argument shows that the K3 surface (\ref{double cover K3 in 3.1.2}) obtained as a double cover of a Halphen surface of index 2 ramified over a singular fiber $k=0$ has a type $I_{2n}$ fiber, $n=4,7,8,9$, when an original Halphen surface has a type $I_n$ fiber as constructed in section \ref{ssec 2.2} and section \ref{ssec 2.3}.

\subsection{Gauge groups in F-theory compactifications on K3 surfaces without a section as double covers of Halphen surfaces with $I_4, I_7, I_8$ fibers}
\label{ssec 3.2}
We discuss the gauge symmetries that arise in F-theory compactifications on the K3 surfaces without a global section that were constructed in section \ref{ssec 3.1} as double covers of Halphen surfaces of index 2 with type $I_4, I_7, I_8$ fibers times a K3 surface. We discuss F-theory compactifications on the K3 surfaces obtained as double covers of Halphen surfaces with a type $I_9$ fiber in section \ref{ssec 3.3} separately. 
\par As described in section \ref{sssec 3.1.1}, K3 surfaces constructed as double covers of Halphen surfaces with a type $I_n$ fiber ramified over a smooth fiber has two type $I_n$ fibers, $n=4,7,8$, and the K3 surfaces have a bisection. Therefore, the gauge symmetries that arise in F-theory compactifications on the K3 surfaces times a K3 surface include a factor as follows:
\begin{equation}
SU(n)^2 \times \Z_2,
\end{equation}
$n=4,7,8$.
\par As described in section \ref{sssec 3.1.2}, K3 surfaces without a global section as double covers of Halphen surfaces with a type $I_n$ fiber, $n=4,7,8$, ramified over a singular fiber at the origin of the base has a type $I_{2n}$ fiber, and the K3 surfaces have a bisection. Therefore, the gauge symmetries arising in F-theory compactifications on the resulting K3 surfaces times a K3 surface have a factor:
\begin{equation}
SU(m) \times \Z_2,
\end{equation}
$m=8,14,16$.

\subsection{Gauge groups in F-theory compactifications on K3 surfaces without a section as double covers of Halphen surfaces with $I_9$ fiber}
\label{ssec 3.3}
We saw in section \ref{ssec 2.4} that the Jacobian fibration of the Halphen surface of index 2 with a type $I_9$ fiber as constructed in section \ref{ssec 2.3} has 1 type $I_9$ fiber and 3 type $I_1$ fibers. Therefore, the Halphen surface with a type $I_9$ fiber as constructed in section \ref{ssec 2.3} has 1 type $I_9$ fiber, 3 type $I_1$ fibers, as well as a double fiber at infinity \footnote{As explained in section \ref{ssec 4.1}, the types of the singular fibers of a Halphen surface of index 2 and the types of the singular fibers of the Jacobian fibration are identical, except the double fiber that a Halphen surface of index 2 possesses.}. Thus, a K3 surface obtained as a double cover of the Halphen surface with type $I_9$ fiber has 2 type $I_9$ fibers and 6 type $I_1$ fibers. We deduce that the non-Abelian gauge group that arises in F-theory compactification on this K3 surface times a K3 surface is precisely
\begin{equation}
SU(9)^2.
\end{equation}
As discussed in \cite{K}, F-theory compactification on a space constructed as the direct product of K3 surfaces yields a four-dimensional theory with $N=2$ supersymmetry, and the anomaly cancellation condition requires that 24 7-branes should be present. Type $I_n$ fiber corresponds to $n$ 7-branes \footnote{The number of 7-branes wrapped on a discriminant component is given by the Euler number of the fiber type over that component. The Euler numbers of the types of the singular fibers of an elliptic surface are given in \cite{Kod2}.}. The number of 7-branes associated with 2 type $I_9$ fibers and 6 type $I_1$ fibers is 24. Therefore, we confirm that the anomaly cancellation condition is satisfied for F-theory compactification on the K3 surface obtained as a double cover of the Halphen surface with a type $I_9$ fiber ramified along a smooth fiber times a K3 surface. A K3 surface obtained as a double cover of the Halphen surface with a type $I_9$ fiber ramified along a smooth fiber has a bisection, therefore the gauge group in F-theory compactification includes the following factor
\begin{equation}
SU(9)^2 \times \Z_2.
\end{equation}
\par Next, we discuss the K3 surface obtained as a double cover of the Halphen surface with a type $I_9$ fiber ramified along a singular fiber. The corresponding quadratic base change ramifies over the type $I_9$ fiber; therefore, the resulting K3 surface has 1 type $I_{18}$ fiber and 6 type $I_1$ fibers. We confirm that this agrees with the anomaly cancellation condition. The non-Abelian gauge group that arises in F-theory compactification on the resulting K3 surface times a K3 surface is precisely given by
\begin{equation}
SU(18).
\end{equation}
The resulting K3 surface has a bisection, thus the arising gauge group contains a factor as follows:
\begin{equation}
SU(18) \times \Z_2.
\end{equation}

\section{Jacobian fibrations of K3 surfaces as double covers of Halphen surfaces of index 2 and $U(1)$ gauge symmetries}
\label{sec 4}
\subsection{General theory of the Jacobian of Halphen surfaces of index 2 and K3 surfaces as double covers of Halphen surfaces}
\label{ssec 4.1}
Halphen surface of index 2 always has the Jacobian fibration. The types of the singular fibers and their locations over the base of a Halphen surface and those of the Jacobian fibration are identical, except the multiple fiber of a Halphen surface. By taking the Jacobian fibration, the double fiber of a Halphen surface becomes a smooth fiber, as mentioned in \cite{Kimura1801}.
\par Halphen surfaces of index 2 and genus-one fibered K3 surfaces without a section as constructed in section \ref{ssec 3.1} are bisection geometries. Therefore, as discussed in \cite{BM}, double covers of quartic polynomials describe these surfaces. Taking the resolvent cubic of quartic polynomials and equating with the term $y^2$ yield the Weierstrass equations of their Jacobian fibrations \cite{BM}.

\subsection{Jacobian fibrations of K3 surfaces as double covers of Halphen surface with a type $I_9$ fiber, and $U(1)$ gauge symmetries in F-theory compactifications}
\label{ssec4.2}
Taking double covers of the Halphen surface with a type $I_9$ fiber as constructed in section \ref{ssec 2.3} yields genus-one fibered K3 surfaces without a section as obtained in section \ref{ssec 3.1}. We deduce the Weierstrass equations of the Jacobians of these K3 surfaces when the ramifications occur along a smooth fiber. We also determine the Mordell--Weil rank of the Jacobians. Using this, we find that F-theory compactifications on generic members of the genus-one fibered K3 surfaces constructed as double covers of the Halphen surface with a type $I_9$ fiber ramified over a smooth fiber do not have a $U(1)$ gauge symmetry.
\par First, we briefly review the quadratic base change \footnote{A mathematical discussion of the quadratic base change can be found in \cite{SchShio}.}. The quadratic base change of an elliptic fibration is an operation in which the coordinate of the base curve $\P^1$ is replaced by homogeneous quadratic polynomial. We denote the coordinate of the base $\P^1$ by $[u:v]$. The quadratic base change of a rational elliptic surface yields an elliptic K3 surface. The process of gluing a pair of identical rational elliptic surfaces with a section to yield an elliptic K3 surface is described by the quadratic base change of the rational elliptic surface, and this can be seen as the reverse of the stable degeneration in which a K3 surface splits into a pair of identical rational elliptic surfaces, as discussed in \cite{Kimura1710}.
\par When the Weierstrass equation of a rational elliptic surface with a section is given as follows:
\begin{equation}
\label{RES in 4.2}
y^2=x^3+f(u,v)\, x+g(u,v),
\end{equation}
we consider the following replacements of the coordinates:
\begin{eqnarray}
\label{base change in 4.2}
u \rightarrow \alpha_1 u^2 +\alpha_2 uv + \alpha_3 v^2 \\ \nonumber
v \rightarrow \alpha_4 u^2 +\alpha_5 uv + \alpha_6 v^2.
\end{eqnarray}
Constants $\alpha_i$, $i=1, \cdots, 6$, in (\ref{base change in 4.2}) give the parameters. These replacements transform the homogeneous polynomial $f$ of degree 4 and the homogeneous polynomial $g$ of degree 6. We denote the resulting homogeneous polynomials by $\til{f}$ and $\til{g}$ which have degrees 8 and 12, respectively. The Weierstrass equation 
\begin{equation}
y^2=x^3+\til{f}\, x+\til{g}
\end{equation}
gives an elliptic K3 surface that the quadratic base change of the original rational elliptic surface (\ref{RES in 4.2}) yields. 
\par As discussed previously in section \ref{sssec 3.1.1}, a K3 surface constructed as a double cover of the Halphen surface with a type $I_9$ fiber is described by the quadratic base change of that Halphen surface. Therefore, the Jacobian fibration of the K3 surface constructed as a double cover of the Halphen surface with a type $I_9$ fiber is given as the quadratic base change of the Jacobian fibration of the Halphen surface with a type $I_9$ fiber, $X_{[9, 1, 1, 1]}$. The Weierstrass equation of the Jacobian fibration of the Halphen surface with a type $I_9$ fiber, $X_{[9, 1, 1, 1]}$, is given as follows \cite{MP}:
\begin{equation}
\label{RES 9111 in 4.2}
y^2=x^3-3u(u^3+24v^3)\, x+2(u^6+36u^3v^3+216v^6).
\end{equation}
The type $I_9$ fiber of the extremal rational elliptic surface $X_{[9, 1, 1, 1]}$ (\ref{RES 9111 in 4.2}) is located over $[u:v]=[1:0]$ in the base $\P^1$ \cite{MP}. Utilizing the quadratic base change (\ref{base change in 4.2}), we find that the Jacobian fibration of a K3 surface constructed as a double cover of the Halphen surface with a type $I_9$ fiber is described by the following Weierstrass equation:
\begin{equation}
\label{JacobianK3 in 4.2}
\begin{split}
y^2= & x^3-3(\alpha_1 u^2 +\alpha_2 uv + \alpha_3 v^2)\cdot [( \alpha_1 u^2 +\alpha_2 uv + \alpha_3 v^2)^3+24(\alpha_4 u^2 +\alpha_5 uv + \alpha_6 v^2)^3]\, x \\
& +2\, [(\alpha_1 u^2 +\alpha_2 uv + \alpha_3 v^2)^6+36(\alpha_1 u^2 +\alpha_2 uv + \alpha_3 v^2)^3\, (\alpha_4 u^2 +\alpha_5 uv + \alpha_6 v^2)^3 \\
& +216(\alpha_4 u^2 +\alpha_5 uv + \alpha_6 v^2)^6].
\end{split}
\end{equation}
\par The Jacobian fibration of a K3 surface constructed as a double cover of the Halphen surface with a type $I_9$ fiber (\ref{JacobianK3 in 4.2}) can be seen as obtained by gluing a pair of identical rational elliptic surfaces (\ref{RES 9111 in 4.2}) \cite{Kimura1710}. Thus, as shown in \cite{Kimura1802}, the Mordell--Weil rank of the Jacobian K3 (\ref{JacobianK3 in 4.2}) is equal to the Mordell--Weil rank of the original rational elliptic surface (\ref{RES 9111 in 4.2}) for generic values of the parameters $\alpha_i$, $i=1, \cdots, 6$. The rational elliptic surface (\ref{RES 9111 in 4.2}) is an extremal rational elliptic surface with the singularity type $A_8$. Therefore, it has the Mordell--Weil rank 0. Thus, it follows that the resulting Jacobian K3 (\ref{JacobianK3 in 4.2}) has the Mordell--Weil rank 0 for generic values of the parameters $\alpha_i$, $i=1, \cdots, 6$. 
\par From this, we conclude that F-theory compactification on a K3 surface constructed as a double cover of the Halphen surface with a type $I_9$ fiber ramified along a smooth fiber times a K3 surface, generically does not have a $U(1)$ gauge symmetry.
\par We also discuss the Jacobian fibration of a K3 surface constructed as a double cover of the Halphen surface with a type $I_9$ fiber ramified along a singular fiber at the origin of the base. We saw in section \ref{ssec 3.3} that a K3 surface constructed as a double cover of the Halphen surface with a type $I_9$ fiber ramified along a singular fiber has 1 type $I_{18}$ fiber and 6 type $I_1$ fibers. Thus, this K3 surface without a section has the singularity type 
\begin{equation}
A_{17}.
\end{equation}
The Jacobian fibration has the identical singularity type $A_{17}$. The complex structure moduli of the elliptic K3 surfaces with a section with the singularity type $A_{17}$ is constructed in \cite{KimuraMizoguchi} by considering a special limit of the quadratic base change of the extremal rational elliptic surface $X_{[9, 1, 1, 1]}$. The Jacobian fibration of the K3 surface obtained as a double cover of the Halphen surface with a type $I_9$ fiber ramified along a singular fiber belongs to this moduli. The members of this moduli, namely elliptic K3 surfaces with a section with the singularity type $A_{17}$, the Weierstrass equations of which are deduced in \cite{KimuraMizoguchi}, correspond to special limits of the base change (\ref{base change in 4.2}) in which the parameters $\alpha_i$ take special values as follows:
\begin{equation}
\alpha_4=\alpha_5=0, \hspace{3mm} \alpha_1\ne 0, \hspace{3mm} \alpha_6\ne 0.
\end{equation}

\section{Conclusions}
\label{sec 5}
We constructed several Halphen surfaces of index 2 with type $I_n$ fibers ($n=4,7,8,9$) in this study. We obtained genus-one fibered K3 surfaces lacking a section by taking double covers of the constructed Halphen surfaces. Two types of K3 surfaces were obtained, depending on whether the ramification locus of a double cover is a smooth fiber, or a singular fiber. These constructions yielded K3 surfaces singular fibers of which include two type $I_n$ fibers, and K3 surfaces singular fibers of which include a type $I_{2n}$ fiber. 
\par We analyzed F-theory compactifications on these K3 surfaces lacking a section times a K3 surface. These K3 surfaces have bisection geometries. Gauge group that arises contains a factor $SU(n)^2 \times \Z_2$ in F-theory compactification on a K3 surface obtained as a double cover of a Halphen surface with a type $I_n$ fiber ramified along a smooth fiber times a K3 surface. Gauge group has a factor $SU(2n) \times \Z_2$ in F-theory compactification on a K3 surface obtained as a double cover of a Halphen surface with a type $I_n$ fiber ramified over a singular fiber times a K3 surface. 
\par We showed that a Halphen surface of index 2 can have a type $I_n$ fiber up to $I_9$ fiber. We constructed a Halphen surface of index 2 with a type $I_9$ fiber, that saturates this upper bound. We determined the complex structure of the Jacobian fibration of this Halphen surface. We precisely obtained the non-Abelian gauge symmetries that arise in F-theory compactifications on K3 surfaces obtained as double covers of the constructed Halphen surface of index 2 with a type $I_9$ fiber times a K3 surface. 
\par Fibering K3 surfaces obtained in this note over a base complex surface can yield genus-one fibered Calabi--Yau 4-folds. An investigation of this construction, and the effect of F-theory compactification on the resulting Calabi--Yau geometries is a likely direction of future studies.

\section*{Acknowledgments}

We would like to thank Shun'ya Mizoguchi and Shigeru Mukai for discussions. This work is partially supported by Grant-in-Aid for Scientific Research {\#}16K05337 from the Ministry of Education, Culture, Sports, Science and Technology of Japan.

\end{document}